\documentclass[aps,preprint,showpacs]{revtex4}%
\usepackage{amsfonts}%
\usepackage{amsmath}%
\usepackage{amssymb}%
\usepackage{graphicx}

\begin{document}

\title{Dependence of the liquid-vapor surface tension on the range of interaction: a test of the law of corresponding states}
\author{Patrick Grosfils}
\email{pgrosfi@ulb.ac.be}
\affiliation{Microgravity Research Center, Chimie Physique E.P. CP 165/62,
Universit\'{e} Libre de Bruxelles, Av.F.D.Roosevelt 50, 1050 Brussels,
Belgium.}
\altaffiliation[Also at ]{Center for Nonlinear Phenomena and Complex Systems CP 231, Universit\'{e} Libre de Bruxelles, Blvd. du
Triomphe, 1050 Brussels, Belgium}
\author{James F. Lutsko}
\affiliation{Center for Nonlinear Phenomena and Complex Systems CP 231, Universit\'{e} Libre de Bruxelles, Blvd. du
Triomphe, 1050 Brussels, Belgium}
\email{jlutsko@ulb.ac.be}
\homepage{http://www.lutsko.com}
\pacs{68.03.Cd,65.20.De}

\begin{abstract}
The planar surface tension of coexisting liquid and vapor phases of a fluid of Lennard-Jones atoms is studied as a function of the range of the potential using both Monte Carlo simulations and Density Functional Theory. The interaction range is varied from $r_c^* = 2.5$ to $r_c^* = 6$ and the surface tension is determined for temperatures ranging from $T^* = 0.7$ up to the critical temperature in each case. The results are shown to be consistent with previous studies. The simulation data are well-described by Guggenheim's law of corresponding states but the agreement of the theoretical results  depends on the quality of the bulk equation of state.
\end{abstract}

\date{\today }
\maketitle

\section{Introduction}

One of the most fundamental properties of a fluid is the surface tension at
the liquid-vapor interface. It would seem that such a fundamental property
would be an ideal candidate for study via computer simulation. However, the
determination of the surface tension from simulation turns out to be frought
with difficulties so that even today there is still a substantial amount of
effort directed towards the development of more reliable algorithms and the
refinement of the reported values even for the paradigmatic case of a simple
fluid modeled with the Lennard-Jones interaction\cite{Mecke_JCP_1997}. One of the primary
difficulties is that in all simulations, the potential is truncated at some
finite range and it happens that the surface tension is very sensitive to the
value of the cutoff. For that reason, an important part of the development of
algorithms has focused on the calculation of the corrections needed to get
the infinite-ranged limit from data obtained using a truncated potential (see, e.g., ref. \cite{Mecke_JCP_1997,duque:8611}). This
sensitivity is therefore a nuisance when the goal is to get the infinite range
result, but in other ways it can made useful. In particular, one of the
important reasons to determine the surface tension from simulation is that it
provides a baseline against which theories of inhomogeneous liquids can be
tested\cite{Katsov, Lutsko_JCP_2008}. For this application, the sensitivity of the surface tension to the
range of the potential can be used as a test of the generality of a theory
which was probably motivated in the first place by its agreement with some
existing simulation data. Furthermore, there has recently been a significant
increase in interest in short-ranged potentials in their own right. This is
due to the fact that certain complex fluids, in particular globular proteins,
can, in a first approximation, be modeled as a simple fluid with a very short
ranged interaction\cite{tWF}. It is therefore  interesting to study the properties
of fluids with these kinds of interactions and to test that existing theories
work in this new domain of interest. For these reasons, we present in this
paper a systematic study of the dependence of the surface tension of a
Lennard-Jones fluid as a function of the range of the potential. 

In this paper, we describe the results of Monte Carlo (MC) simulations of a Lennard-Jones fluid 
with the potential truncated at several different points. We have
chosen to truncate and shift the 
Lennard-Jones potential, $v_{LJ}(r)$, so that the potential used in this
work is $v(r;r_c)=v_{LJ}(r)-v_{LJ}(r_c)$ for $r<r_c$ and $v(r)=0$ for
$r \ge r_c$. If this shift is not performed, then there is an impulsive
contribution to the pressure when atoms move across the $r=r_c$
boundary that would have to be taken into account\cite{FrenkelSmit}. We do
not shift the force, i.e. we do \emph{not} use
$\tilde{v}(r;r_c)=v_{LJ}(r)-v_{LJ}(r_c)-(r-r_{c})v_{LJ}'(r_c)$ with
$v_{LJ}'(r)=dv_{LJ}(r)/dr$  inside the cutoff, as is usually done
in molecular dynamics simulations to avoid impulsive forces: our
potential is truncated and shifted but the force is not shifted. This
choice was made in order to allow for comparison with previous MC
studies. 

In the simulations a slab of liquid is 
bounded on both sides by vapor. The surface tension is determined using the method of Bennett\cite{Bennett, FrenkelSmit} as there seems to be
some evidence that this method is more robust than other commonly used
techniques\cite{Errington_JCP_2007}. It is often the case that the
quantity of interest is the surface tension for the infinite-ranged
potential. Since simulations almost always make use of truncated
potentials, various techniques have been developed to approximate the
so-called long range corrections, i.e. the difference between
quantities calculated with the truncated potential and the
infinite-ranged quantities\cite{duque:8611}. We do not include any such corrections
here since our goal is actually to study the truncated
potentials. Thus, each value of the cutoff defines a different
potential with its own coexistence curve and thermodynamics.

In Section II, we
present the simulation techniques used in our work. Section III contains a
discussion of our results including a comparison to previous work.
Since one of the motivations for this work is to provide a baseline for
testing theories of the liquid state, we illustrate this by comparing
our results to Density Functional Theory (DFT) calculations and by
testing the law of corresponding states. We give our conclusions in the
last Section.

\bigskip

\section{Simulation Methods}%
Simulations are performed with a standard Metropolis Monte-Carlo algorithm (MC-NVT) for a system of $N=2000$ particles
of mass $m$ at temperature $T$ in a volume $V=L_xL_yL_z$ where $L_x$, $L_y$, and $L_z$ are the dimensions
of the rectangular simulation cell. Periodic boundary conditions are used in all directions. Particles interact via the
Lennard-Jones potential,
\begin{equation}
v_{LJ}(r)=4\epsilon\left(\left(\frac{\sigma}{r}\right)^{12}-\left(\frac{\sigma}{r}\right)^{6}\right) 
\end{equation}
which is truncated and shifted so that the potential simulated is
\begin{equation}
v(r)\,=\,\left\{\begin{array}
{r@{\quad:\quad}l}
v_{LJ}(r) -\,v_{LJ}(r_c)& r\,<\,r_c\\
0 & r\geq r_c
\end{array}\right.
\end{equation}
where $r_c$ is the cutoff radius. Each simulation starts from a rectangular box ($L_x=L_y=L$, $L_z=4\,L$) filled with a dense disordered arrangement of particles 
($\rho^{\ast} \equiv \rho \sigma^{3} = 0.8$) surrounded along the z-direction by two similar rectangular boxes containing particles in a low density state 
($\rho^{\ast}\sim 0.01$) . The total simulation box has sides of  length $L_x=L_y=9.15\sigma$ and $L_z=109.63\sigma$. The liquid film located in the middle of the box has a thickness $\Delta z\simeq 27\sigma$ so that the two interfaces do not influence each others.  The system is first equilibrated during $5\times 10^5$ Monte-Carlo cycles (one cycle $=\,N$ updates) after which the positions of the particles are saved every $20$ cycles during $5\times 10^5$ cycles. This ensemble of $2.5\times 10^4$ configurations is used to compute the density profile and the surface tension by the Bennett's method.

Although several methods  are available for the computation of the surface tension, the Bennett's approach has been chosen because 
of its accuracy\cite{Errington_JCP_2007}. 
In the Bennett's method the calculation of the surface tension follows from the definition 
\begin{equation}
\gamma\,=\,\left(\frac{\partial F}{\partial A}\right)_{N,V,T}
\end{equation}
where $F$ is the free energy and $A$ is the area of the liquid-vapor interface. In its implementation the method requires that one performs two simulations: one for system $0$ of interface area $A_0$, and another for system $1$ of  interface area $A_1=A_0+\Delta A$. In this work $\Delta A/A=5\times 10^{-4}$. The free energy difference $\Delta F$ between the two systems is evaluated by the method of acceptation ratio which starts with the computation of $\Delta E_{01}=E_{01}-E_{00}$  which is the difference between $E_{00}$, the energy of a configuration of system $0$,  and $E_{01}$, the energy of a new configuration obtained from the previous one by rescaling the positions of the particles\cite{SalomonsMareschal,FrenkelSmit} :
$x'=x\,(A_1/A_0)^\frac{1}{2}$, $y'=y\,(A_1/A_0)^\frac{1}{2}$, and $z'=z\,(A_0/A_1)$. Similarly one computes
$\Delta E_{10}=E_{10}-E_{11}$ obtained from a configuration of system $1$ following an inverse rescaling of the positions.  $\Delta F$ is obtained by requiring that
\begin{equation}
\sum_{n_0}\,f(\Delta E_{01}-\Delta F)\,=\sum_{n_1}\,f(\Delta E_{10}+\Delta F)
\end{equation}
where $\sum_{n_0}$ ($\sum_{n_1}$) is a sum over the configurations of systems $0$ ($1$), and $f(x)=(1+\exp(\beta x))^{-1}$.
Then, taking into account the fact that the system contains two flat interfaces, the value of the surface tension is given by 
$\gamma=\Delta F/(2\Delta A)$ .

\bigskip

\section{Results}%

\subsection{Comparison to previous results}
\begin{figure}[tbp]
\begin{center}
\resizebox{12cm}{!}{
{\includegraphics[height = 12cm, angle=-90]{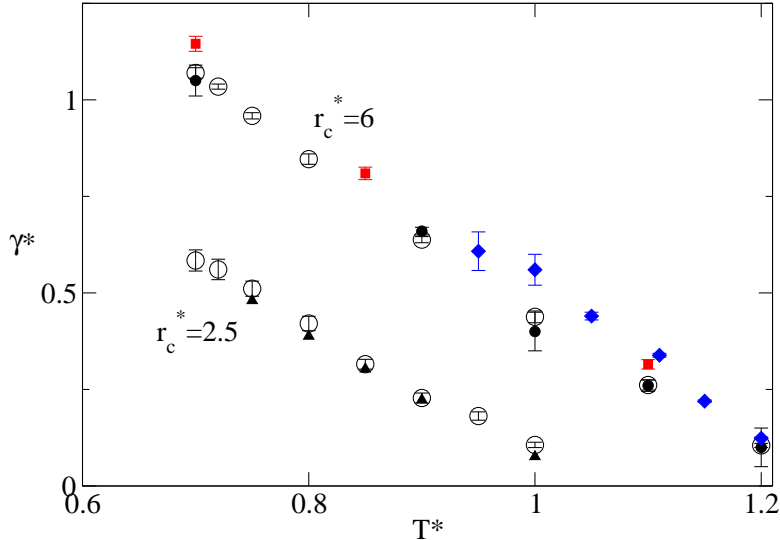}}}
\end{center}
\caption{(Color online)The surface tension as a function of temperature for two different cutoffs. The
open circles are our data, the  filled circles are from Duque et al\cite{Duque_JCP_2004},  
the squares are from Mecke et al\cite{Mecke_JCP_1997}, the diamonds are from Potoff et al\cite{Potoff_JCP_2000} and the triangles are from 
\protect\cite{Haye_Bruin_JCP}). Note that the Mecke and Potoff data both include long-ranged corrections.}
\label{fig1}
\end{figure}

Our results for the surface tension as a function of the cutoff are given in Table 1. Note that all quantities are reported in reduced units so that the reduced temperature is $T^{\ast} = T/ \epsilon$, the reduced cutoffs are $r_{c}^{\ast} = r_{c}/\sigma$ and the reduced surface tension is $\gamma^{\ast} = \gamma \sigma^{2}/\epsilon$. In Fig. \ref{fig1} we show our results for cutoffs of $r_c^*=2.5$ and $6.0$ compared to the MC data of Haye and Bruin\cite{Haye_Bruin_JCP} for the shorter cutoff and to the MD data of Duque et al\cite{Duque_JCP_2004} (who appear to shift the forces) and Potoff et al\cite{Potoff_JCP_2000} and Mecke et al\cite{Mecke_JCP_1997}. The latter two are shown even though they include long-ranged corrections. Our data are seen to be very consistent with the MC data obtained without long-ranged corrections and to lie slightly below the corrected data, as expected. 

\subsection{Comparison to DFT}
\begin{figure}[tbp]
\begin{center}
\resizebox{12cm}{!}{
{\includegraphics[height = 12cm, angle=-90]{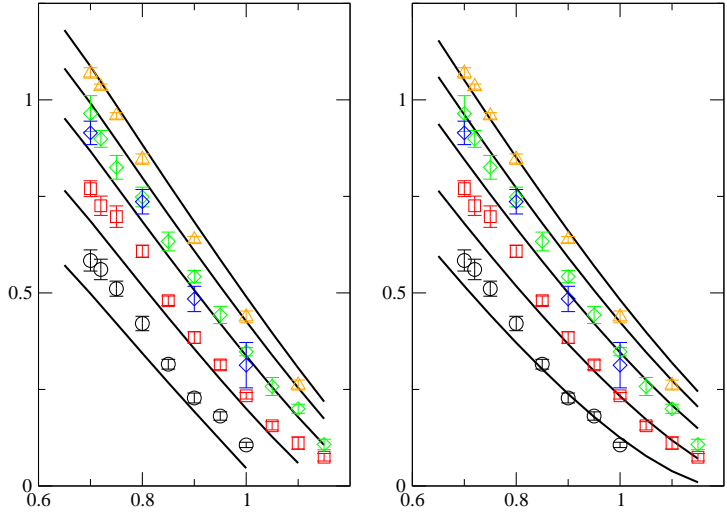}}}
\end{center}
\caption{(Color online)Comparison of our simulation data to a DFT
  model\cite{Lutsko_JCP_2008}. The panel on the left shows the results
  of the theory using an empirical equation of state while the results
  on the right were obtained using thermodynamic perturbation
  theory.The lines are ordered from the smallest cutoff (lowest lines)
  to the largest cutoff (highest lines) and were calculated for
  $r_c^*=2.5,3,4,6,\infty$. The data is represented by circles(2.5),
  squares(3), diamonds(4), filled diamonds (4 - larger system) and triangles(6).}
\label{dft}
\end{figure}

In Fig. \ref{dft}, we compare our results to the predictions of a recently proposed Density Functional Theory model\cite{Lutsko_JCP_2008}. The DFT requires knowledge of the bulk equation of state and the figure shows results using two different inputs: the 33-parameter equation of state of  Johnson, Zollweg and Gubbins (JZG)\cite{JZG} and first order Barker-Henderson thermodynamic perturbation theory\cite%
{BarkerHend,HansenMcdonald}. Both versions of the theory are in good
qualitative agreement with the data, showing the decrease in surface
tension as the range of the potential decreases. It might be thought
that use of an empirical equation of state should automatically give
superior results to an approximation, like thermodynamic perturbation
theory, but this is not necessarily the case since the equation
of state is fitted to data for the infinite-ranged potential. The
finite cutoff is accounted for using simple mean-field
corrections\cite{JZG,FrenkelSmit} and these become increasing
inaccurate as the cutoff becomes shorter and, for fixed cutoff, as the
fluid density becomes higher. The latter condition means, in the
present context, increasing inaccuracy as the temperature
decreases. Both of these trends are confirmed by the figure. The
decrease in accuracy with decreasing cutoff can be seen in the fact
that the critical point (corresponding to the temperature at which the
surface tension extrapolates to zero) is less accurately estimated for
the smaller cutoffs than for the larger cutoffs.

The
perturbation theory, on the other hand, takes the cutoff into account
more accurately and consistently so that no strong change in accuracy
is expected as the cutoff decreases. However, the theory itself is
expected to be less accurate for higher densities so again a drop in
accuracy with decreasing temperature would be expected and that is
indeed seen in the figure. Furthermore, perturbation theory is in
general going to be inaccurate near the critical point as it does not
take into account renormalization effects which tend to lower the
critical point. These effects are less pronounced for shorter-ranged
potentials and indeed the perturbation theory seems more consistent
with shorter-ranged potential.

\begin{table}[tbp]
\caption{Surface tension determined from simulation as a function of temperature for different cutoffs.}
\label{tab1}%
\begin{ruledtabular}
\begin{tabular}{cccccc}
Temperature & $r_c^* = 2.5$\footnotemark[1] & $r_c^* = 3$\footnotemark[1] & $r_c^* = 4$\footnotemark[1] & $r_c^* = 4$\footnotemark[2] & $r_c^* = 6$\footnotemark[2] \\ \hline
0.70 & 0.584 (27) &  0.770 (21) & 0.964(46) & 0.914(30)  & $1.070(13)$  \\
0.72 & 0.561 (26) &  0.726 (25) & 0.899(22) &            &  $1.034(7)$  \\
0.75 & 0.511 (20) &  0.698 (28) & 0.825(31) &            & $0.959(8)$  \\
0.80 & 0.421 (19) &  0.608 (16) & 0.748(25) &  0.736 (32)& $0.847(13)$  \\
0.85 & 0.315 (13) &  0.480 (12) & 0.633(24) &            &   \\
0.90 & 0.228 (13) &  0.384 (15) & 0.542(16) & 0.484(33)  & $0.638(8)$  \\
0.95 & 0.181 (11) &  0.314 (12) & 0.443(22) &            &   \\
1.0  & 0.106 (7) &  0.234 (8)  & 0.348(11) & 0.313 (59) & $0.438(15)$  \\
1.05 & &  0.156 (11) & 0.258(23) &            &   \\
1.1  & &  0.111 (16) & 0.200 (12)&            & $0.261(13)$  \\
1.15 & &  0.074 (10) & 0.108 (14)&            &  \\
1.2  & &  0.054 (6)  & 0.067 (15)& 0.063 (19) & $0.105(5)$  \\
\end{tabular}
\end{ruledtabular}
\footnotetext[1]{using approximately 2000 atoms.}
\footnotetext[2]{using approximately 8000 atoms.}
\end{table}

\subsection{Corresponding states}
\begin{figure}[tbp]
\begin{center}
\resizebox{12cm}{!}{
{\includegraphics[height = 12cm, angle=-90]{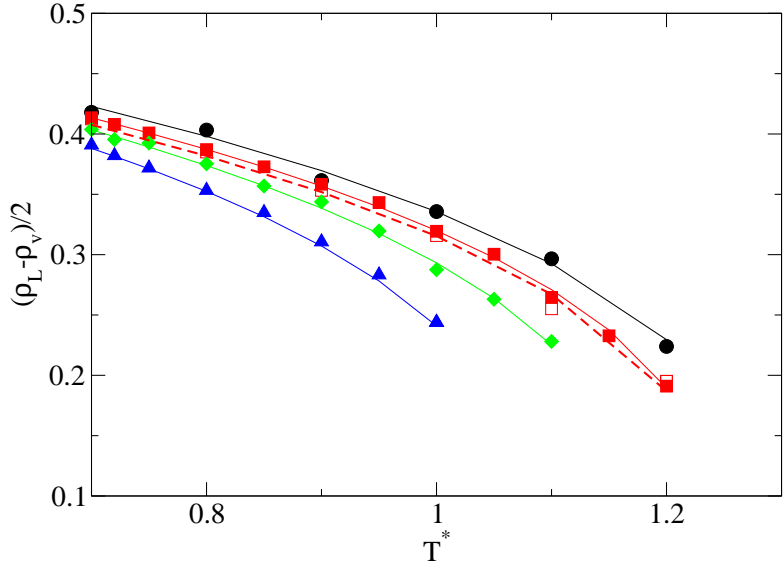}}}
\end{center}
\caption{(Color online) The fit of the difference in liquid and vapor densities, as determined from simulation (symbols), to the RG functional form (lines). The data are shown as circles ($R_c=6.0$), open squares ($R_c=4.0$, larger cell), filled squares ($R_c=4.0$, smaller cell), diamonds ($R_c=3.0$) and triangles ($R_c=2.5$).}
\label{Tc}
\end{figure}

The principle of corresponding states is a generalization of the results of the van der Waals equation of state\cite{Guggenheim}. The idea is that the properties of simple liquids should be universal functions of the state variables, density and temperature, scaled to the critical point. In this section, we test this hypothesis by applying it to the surface tension. The first step is therefore to determine the critical temperatures and densities of the various truncated potentials. Since the theoretical calculations require as input an equation of state, the critical points are easily determined. To determine them from the simulations, we took five independent averages over 5000 configurations and fitted the density profiles in each case to a hyperbolic tangent and then from these extract the coexisting vapor and liquid densities at each temperature. The five values obtained at each temperature were averaged and the variance used as an estimate of the errors in the values. The critical temperature was then estimated by using the lowest order renormalization group (RG) result, $\frac{1}{2}(\rho_l-\rho_v)=A(T_c-T)^{0.325}$\cite{Zinn1,Zinn2}. In some applications\cite{LiquidVaporCurves}, higher order terms are included but we did not feel that the accuracy of our data warranted use of anything but the lowest order function. The critical density was then estimated using the law of the rectilinear diameter, $\frac{1}{2}(\rho_l+\rho_v)=\rho_c+B(T_c-T)$\cite{Guggenheim}. There are again higher order corrections to this formula which can be calculated using RG methods, but for the reasons just given, we have not attempted to include them. The results of these fits are illustrated in Fig. \ref{Tc} and summarized in Table 2. The largest errors in this procedure are in the determination of the critical density.

\begin{table}[tbp]
\caption{The critical points for the LJ potential truncated at different values. The theoretical values were determined using the empirical JZG equation of state\cite{JZG} (JZG) and the first order Barker-Henderson perturbation theory (BH).}
\label{tab2}%
\begin{ruledtabular}
\begin{tabular}{c|cc|cc|cc}
\multicolumn{1}{c|}{ $R^*_c$} & \multicolumn{2}{c|}{MC} & \multicolumn{2}{c}{Theory - JZG}  & \multicolumn{2}{c}{Theory - BH} \\  
& $T^*_c$ & $\rho^*_c$ & $T^*_c$ & $\rho^*_c$ & $T^*_c$ & $\rho^*_c$ \\ \hline
2.5 & 1.10 (1)\footnotemark[1] &  0.31 (9)\footnotemark[1] & 1.04 & 0.26 & 1.18 & 0.325\\
3.0 & 1.18 (1)\footnotemark[1] &  0.31 (7)\footnotemark[1] & 1.15 & 0.28 & 1.27 & 0.342 \\
4.0 & 1.26 (1)\footnotemark[1] &  0.31 (5)\footnotemark[1] & 1.25 & 0.32 & 1.34 & 0.341 \\
4.0 & 1.25 (2)\footnotemark[2] &  0.30 (6)\footnotemark[2] & --- & --- & --- & --- \\
6.0 & 1.30 (2)\footnotemark[2] &  0.32 (9)\footnotemark[2] & 1.29 & 0.35 & 1.38 & 0.341 \\
$\infty$ & 1.31 \footnotemark[3] & 0.317 \footnotemark[3]  & 1.311 & 0.351 & 1.40 & 0.312 \\
\end{tabular}
\end{ruledtabular}
\footnotetext[1]{using approximately 2000 atoms.}
\footnotetext[2]{using approximately 8000 atoms.}
\footnotetext[3]{From ref. \cite{perez-pellitero:054515}.}
\end{table}

\begin{figure}[tbp]
\begin{center}
\resizebox{12cm}{!}{
{\includegraphics[height = 12cm, angle=-90]{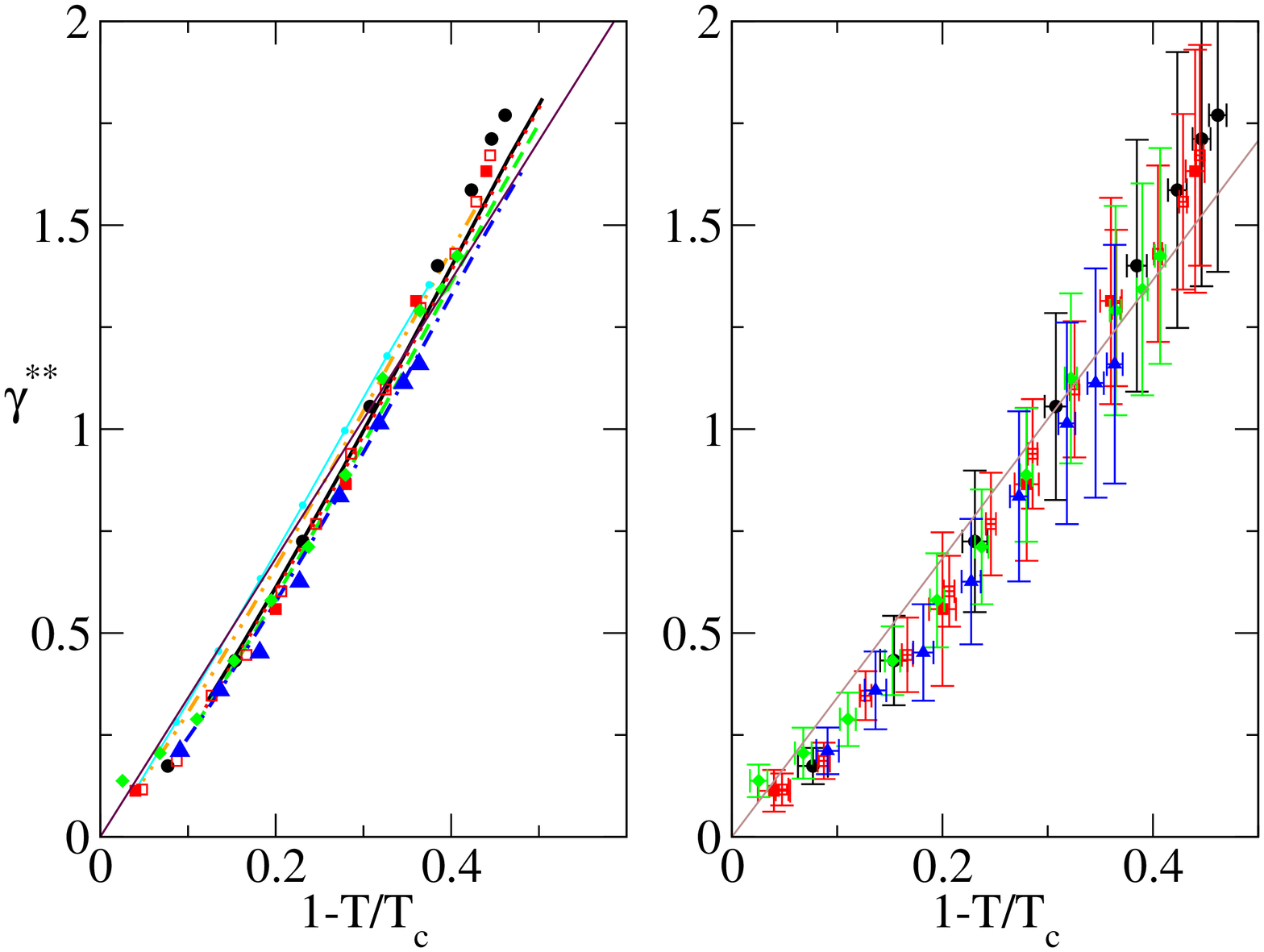}}}
\end{center}
\caption{(Color online) The scaled surface tension, $\gamma^{**} \equiv \frac{\gamma}{T_c \rho_c^{2/3}} $, as a function of distance from the critical temperature. The left panel includes the theoretical curves, shown as full line ($R^*_c=\infty$), dotted line ($R_c^*=8$), dashed line ($R_c^*=6$), dash-dot line ($R_c^*=4$), dash-dot-dot line ($R_c^*=3$), and line+circles ($R_c^*=2.5$), and the simulation data, shown as circles ($R_c^*=6$), filled squares ($R_c^*=4$, 2000 atoms), open squares ($R_c^*=4$, 8000 atoms), diamonds ($R_c^*=3$) and triangles ($R_c^*=2.5$). The right hand panel shows only the data from simulation as well as the estimated error. In both cases, the thin line is a best fit to all of the data (theory and simulation) of the form $\gamma^*=\gamma^*_0(1-T/T_c)$ with $\gamma^*_0=3.41$.}
\label{corresponding}
\end{figure}

\begin{figure}[tbp]
\begin{center}
\resizebox{12cm}{!}{
{\includegraphics[height = 12cm, angle=-90]{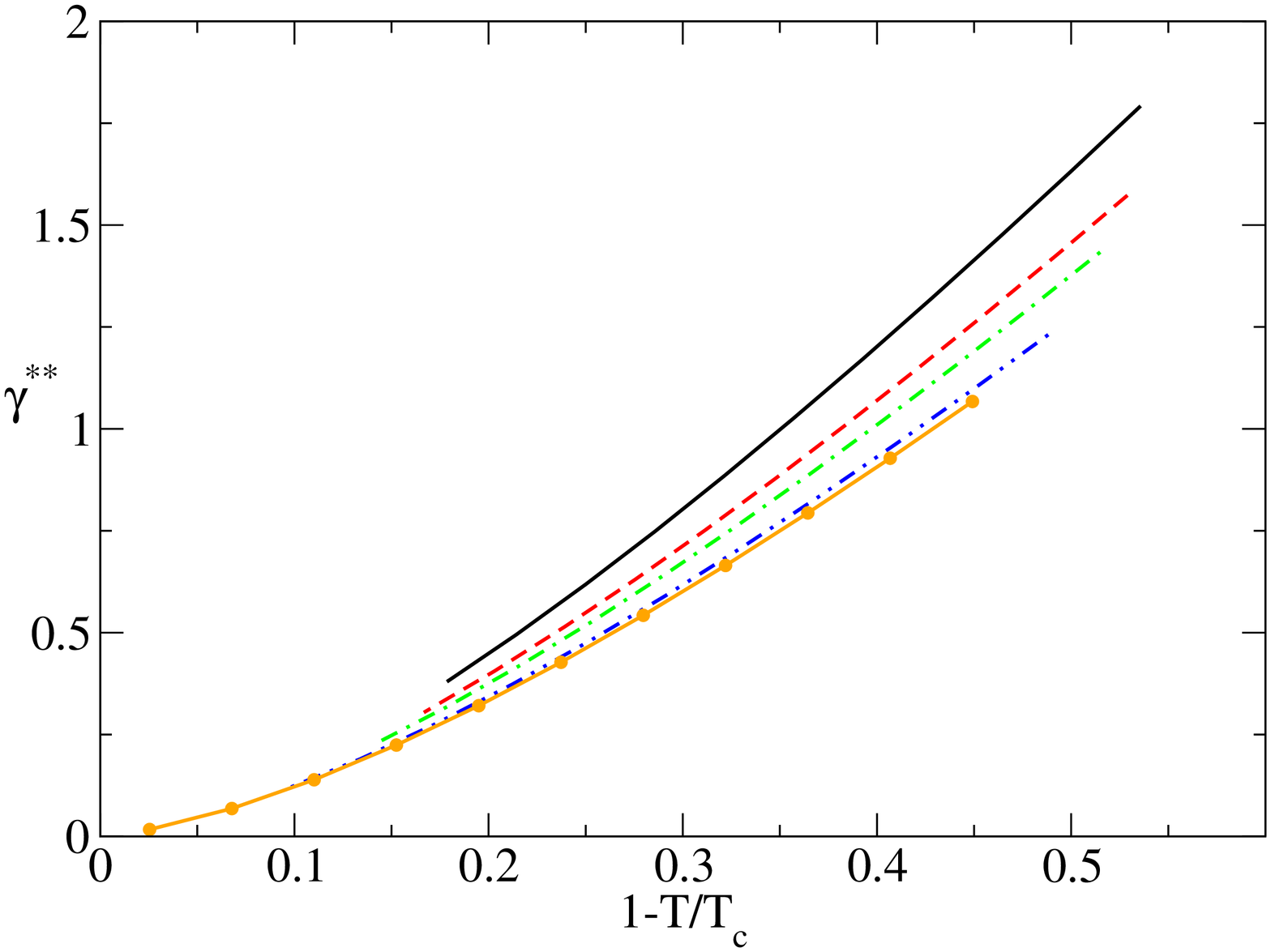}}}
\end{center}
\caption{(Color online) The scaled surface tension, $\gamma^{**} \equiv \frac{\gamma}{T_c \rho_c^{2/3}} $, as a function of distance from the critical temperature as calculated using the perturbative equation of state displayed as a full line ($R^*_c=\infty$), dashed line ($R_c^*=6$), dash-dot line ($R_c^*=4$), dash-dot-dot line ($R_c^*=3$), and line+circles ($R_c^*=2.5$).}
\label{bh}
\end{figure}

Figure \ref{corresponding} shows the surface tensions, as determined from simulation and theory using the JZG equation of state, scaled to the critical density and temperature as a function of distance from the critical temperature. Despite the wide range of cutoffs and the mixture of data from simulations and theory, it is nevertheless seen that the data do in fact obey the law of corresponding states to a good approximation. However, the same scaling of the theoretical calculations using the equation of state from thermodynamic perturbation theory, shown in Figure \ref{bh}, does not give a single curve. While the data for the shorter cutoffs, $r_c^*=2.5$ and $r_c^*=3$ appear to coincide, the data for the larger cutoffs does not. This appears to be due, at least in part, to the fact that the estimate of the critical density as a function of the cutoff calculated using the perturbation theory is not monotonic (see Table \ref{tab2}) which is at odds with the quantities as determined from simulation which clearly \emph{are} monotonic in the cutoff. 

\bigskip

\section{Conclusions}
We have presented our determination of the liquid-vapor surface
tension in a Lennard-Jones fluid as a function of the range of the
potential. The data give a systematic picture of the variation of surface tension with the cutoff
 and are in agreement with previous studies. It is hoped that this can serve as a useful benchmark for
the development of theories of inhomogeneous liquids. Indeed, the
results were compared here to calculations made using a recently
developed DFT and the strengths and weaknesses of the theory are
evident: while it gives a good semi-quantitative estimate of the
surface tension for all cutoffs, errors on the order of 10\% are
present indicating that further improvement is possible.  

We have also tested the law of corresponding states by showing our results from both simulation and theory
scaled to the critical density and temperature. For the simulation data and the theoretical
calculations based on an empirical equation of state, the law of corresponding states
appears to be obeyed. However, the calculations based on the equation of state
from first order perturbation theory do not appear to scale well at all. This failure appears to be
due to poor behavior of the critical density as a function of the cutoff and indicates that care must be
exercised before using the law of corresponding states to extrapolate calculations.

\begin{acknowledgments}
This work was supported by the European Space Agency under contract
number ESA AO-2004-070 and by the projet ARCHIMEDES of the Communaut\'e
Fran\c caise de Belgique (ARC 2004-09). 
\end{acknowledgments}

\bibliographystyle{apsrev}
\bibliography{mc-surfacetension}

\end{document}